\documentstyle[amsmath,aps,prb,multicol,epsf,floats,axodraw,amssymb,eqsecnum,graphics]{revtex}
\def\alb{{\alpha\beta}}
\def\non{\nonumber \\}
\begin{document}
\draft
%***********    This is for two columns *******************************
\twocolumn[\hsize\textwidth\columnwidth\hsize\csname @twocolumnfalse\endcsname
%********************************
\title{Order induced by dipolar interactions in a geometrically
frustrated antiferromagnet}
\author{S. E. Palmer and J. T. Chalker}
\address{Theoretical Physics, University of Oxford, 1 Keble Road,
Oxford  OX1 3NP, UK}
\maketitle
\date{\today}
\maketitle
\begin{abstract}
We study the classical Heisenberg model for 
spins on a pyrochlore lattice interacting via long range
dipole-dipole forces and nearest neighbor exchange. 
Antiferromagnetic exchange alone is known not to induce ordering 
in this system.
We analyze low
temperature order resulting from the combined interactions, 
both by using a mean-field approach
and by examining the energy cost of fluctuations about an ordered
state.  We discuss behavior as a function of the ratio of the dipolar
and exchange interaction strengths and find two types of ordered phase.
We relate our results to the recent experimental work and reproduce
and extend the theoretical calculations on the pyrochlore compound,
Gd$_2$Ti$_2$O$_7$, by Raju \textit{et al.}, Phys. Rev. B {\bf 59},
14489 (1999).
\end{abstract}
\vskip 0.2 truein
\newpage
% ***********    This is for two columns *******************************
\vskip2pc]
%********************************
\section*{Introduction}
Geometrically frustrated magnetic systems have received a great deal
of attention in recent years from both experimentalists and
theorists.\cite{reviews} Such systems are typically composed of corner- or 
edge-sharing  frustrated units, usually triangles or tetrahedra.  
For example, a
two-dimensional network of corner-sharing triangles forms the
kagom\'{e} lattice while the three-dimensional pyrochlore structure is
composed of corner-sharing tetrahedra.  Magnetic systems with such
structure exhibit unusual low temperature properties which are not
completely understood.  These materials typically remain disordered to
a freezing temperature $T_g \ll
|\theta_{CW}|$.\cite{reviews,ginfreeze} Indeed, the depression of the
transition temperature below that expected from the measured value
of $|\theta _{CW}|$ is often used to gauge the degree of frustration
in a magnetic system.

The reluctance of geometrically frustrated antiferromagnets to order
can be understood by considering classical models, many examples of
which yield macroscopically degenerate ground states.  For instance,
it has been shown that the Heisenberg model for spins on a pyrochlore
lattice interacting via nearest neighbor antiferromagnetic exchange
has an extensive number of degrees of freedom in the ground
state.\cite{villain79,reimers91,reimers92,moeschalk} The degeneracy of the
ground state manifold is not lifted in this model by thermal
fluctuations and the system has no finite temperature ordering
transition.\cite{reimers92,moeschalk} Several realizations of
pyrochlore magnets have been studied
experimentally.\cite{ginfreeze,zinkin97,spinice,dipolar}

Classical models which include nearest neighbor exchange interactions
alone may not be sufficient for explaining all the low temperature
properties of some frustrated antiferromagnets.  For instance, the
origin of the freezing transition must involve additional aspects of
the system.  In real materials, several other types of interactions
are present.  Further neighbor exchange may be
relevant\cite{reimers91} as may crystal field
effects\cite{bramwell94}, magnetic dipole
interactions\cite{dipolar} or the effect of quenched
disorder\cite{villain79,disorder}.

In this paper we focus on how the inclusion of dipolar forces affects
the properties of the pyrochlore antiferromagnet.  Because of the
ground state degeneracy of this system with only nearest neighbor
exchange, dipolar interactions are important in establishing order even
if they are weak.  Moreover, the influence of dipole interactions in a
pyrochlore antiferromagnet has been probed\cite{raju99} in the
compound Gd$_2$Ti$_2$O$_7$, which should be well represented by an
isotropic Heisenberg model.

The properties of Gd$_2$Ti$_2$O$_7$ have been studied both
experimentally and theoretically in a recent paper by Raju \textit{et
al.}, in which it was shown that Gd$_2$Ti$_2$O$_7$ undergoes a
transition to long range order at a temperature of about
1K.\cite{raju99} Their measurement of the high temperature
susceptibility gives a negative Curie--Weiss constant,
$\theta_{CW}\simeq -9.8$K.\cite{raju99} Measurements on a magnetically
dilute sample show a reduction in $|\theta_{CW}|$, indicating that
this value is predominantly due to antiferromagnetic
exchange.\cite{raju99} The transition temperature is much lower than
$|\theta_{CW}|$, indicating that the system is frustrated.  The
theoretical work of these authors involves mean-field calculations
expressed as a Landau expansion of the free energy and taken to
quadratic order.  They examine the ordering instabilities that occur
as the temperature is lowered.  With nearest neighbor exchange and
long-range dipolar interactions, they find that order parameter
fluctuations on entire branches in $q$-space along the star of the
$(111)$ direction become unstable simultaneously at
$T_c$.\cite{raju99} Degeneracies of this kind are often broken by
thermal or quantum fluctuations, a phenomenon known as
order-by-disorder.\cite{obd} Raju \textit{et al.}  suggest that this
mechanism may operate to induce order in the model they
study.\cite{raju99}
%Alternatively, by considering further
%neighbor interactions of various sign, i.e. combinations of
%ferromagnetic and antiferromagnetic exchange, they find that different
%individual incommensurate modes are selected out depending on the
%relative strength of the second or third neighbor exchange as compared
%to the dipole-dipole coupling.\cite{raju99}

In the following, we extend this mean-field description of the system
with only nearest neighbor exchange and long range dipolar
interactions to find the ordering pattern below $T_c$.  We show that
the quartic term in the free energy expansion lifts the degeneracy of
the critical modes.  The ordering pattern obtained for the ratio of
dipolar to exchange interaction strengths appropriate for
Gd$_2$Ti$_2$O$_7$ is a four-sublattice N\'{e}el state.  In addition,
we analyze the low temperature fluctuations away from the four
sub-lattice ground state.  We show that all distortions have a
positive energy cost.  Ordering in this model, therefore, has an
energetic origin and is not an example of fluctuation-induced order.

\section{Spins on a single tetrahedron}
\begin{figure}
\begin{center}
\leavevmode
\hbox{%
\epsfysize=2in
\epsffile{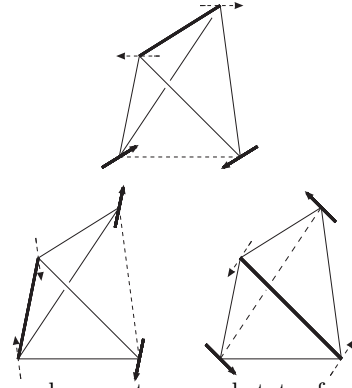}}
\caption{Three degenerate ground states for spins on a single
tetrahedron.  The spins in each configuration are parallel to certain
edges of the tetrahedron.  These edges and the spins parallel to them
are drawn with the same type of line, either bold or
dashed. There are three other ground states, 
obtained from these by reversing all spins.\label{gnds}}
\end{center}
\end{figure}

It is instructive first to consider spins on a single tetrahedron.
Labeling the spins by $i$,$j$, the interaction
energy for the tetrahedron is
\begin{align}
U_{\mathit{int}}=&\frac{J_{ex}}{2}\sum_{i\neq j}{\bf S}_i\cdot{\bf S}_j\non
&+J_{dd}\cdot \frac{A}{2}\sum_{i\neq j}[{\bf S}_i\cdot{\bf S}_j
-3({\bf S}_i\cdot\hat {\bf r}_{ij})({\bf S}_j\cdot\hat {\bf r}_{ij})]
\end{align}
where $J_{ex}$ and $J_{dd}$ are the relative interaction strengths of
the exchange and dipolar terms, $A=\mu^2/{a^3}$, $a$ is the edge
length of the tetrahedron, and $\mu$ is the dipole moment.  We
minimize this energy using a standard numerical search and find all
the possible ground states.  These are shown in Fig.~\ref{gnds}.  In
each of the ground state configurations, the spins are coplanar and
are tangent to the sphere circumscribing the tetrahedron.  It is
worthwhile to note that these states are also ground states for spins
interacting via nearest neighbor antiferromagnetic exchange only.
Without dipolar interactions, the condition for a configuration to be
a ground state is that the vector sum of the spins on each tetrahedron
is zero, leaving two internal degrees of freedom for the
configuration.\cite{reimers91,moeschalk} Dipolar interactions fix
these two degrees of freedom and those arising from the ${\mathcal
O}(3)$ symmetry.

Since the interaction energy for a single tetrahedron involves only 
nearest-neighbor dipolar contributions, one might not
expect the spin configurations in Fig.~\ref{gnds} to be a useful
guide to behavior on the full pyrochlore lattice. In fact,
and somewhat surprisingly, the ground state for the full lattice that we
find in Sec.\,III turns out to be a periodic repetition of that for
a single tetrahedron, provided the ratio of dipolar to exchange interactions 
does not exceed a critical value.

\section{Mean-field theory for interactions on the pyrochlore lattice}
We consider a system of spins interacting via the Hamiltonian
\begin{equation*}
{\mathcal H}=\frac{1}{2}\sum_{i\neq j,\alb}S_i^\alpha
J_{ij}^{\alb}S_j^\beta 
\end{equation*}
with
\begin{equation}
J^{\alb}_{ij}=\delta_{\alb}\delta_{nn}+\frac{C}{r_{ij}^3}
(\delta_{\alb}-\frac{3(r_{ij}^\alpha)(r_{ij}^\beta)}{r_{ij}^2})
\end{equation}
where $C=\mu^2J_{dd}/J_{ex}$. The exchange interaction is between
nearest neighbors only, as indicated by $\delta_{nn}$.  The labels $i$
and $j$ refer to sites in the lattice and $\alpha$ and $\beta$ label
components of the spin or spatial vectors.  A mean-field treatment for
Hamiltonians of this kind has been developed by Reimers, Berlinsky and
Shi.\cite{reimers91} For completeness, we briefly summarize their
approach.  Order parameters for the system are defined by
\begin{equation}
{\rm Tr}(\rho_i {\bf S}_i)={\bf B}_i
\end{equation}
where $\rho_i$ is the local density matrix and is constrained by ${\rm
Tr}(\rho_i)=1$.  Expanding the free energy in a power series in ${\bf
B}_i$ yields
\begin{eqnarray}
%F&=&4NT\ln\left[\frac{\Gamma(\frac{3}{2})}{2\pi^{3/2}}\right]\non
F&=&\mathrm{const.}\qquad\qquad\non
&&+\frac{1}{2}\sum_{i\neq j, \alb}B_i^\alpha (3T\delta_\alb \delta_{ij}
+J_{ij}^\alb) B_j^\beta\non 
&&+\frac{9}{20}T\sum_{i,\alb}(B_i^\alpha
B_i^\alpha)(B_i^\beta B_i^\beta)+{\mathcal O}(B^6). 
%-\sum_{i,\alpha}H^\alpha B_i^\alpha
\end{eqnarray}
The next step is to diagonalize the quadratic term.  In systems with
$n$ atoms per unit cell, it is convenient to divide the site label $i$
into two parts: $l$ labels the unit
cell and $a$ labels spins within the unit cell.  Making the Fourier
transforms
\begin{eqnarray}
B_{la}^\alpha&=&\sum_{\bf q}B_{\bf q}^{a\alpha}e^{i{\bf q}\cdot{\bf r}_l}\\
J_{lmab}^{\alpha\beta}&=&\frac{1}{N}\sum_{\bf q}J_{\bf q}^{a\alpha b\beta}e^{-i{\bf
q}\cdot{\bf r}_{lm}}
\end{eqnarray}
and substituting these expressions into the free energy, one arrives at the
expression for the quadratic part of $f=F/N$, the free energy per
unit cell, used in Ref.\,\onlinecite{raju99}:
\begin{equation}
f^{(2)}=\frac{1}{2}\sum_{ab,\alb}\sum_{\bf q}B_{\bf q}^{a\alpha}
\left(3T\delta_{ab}\delta_{\alb}+J_{\bf q}^{a\alpha b\beta}\right)B_{-{\bf q}}^{b\beta}.
%&&+\frac{9T}{20}\sum_{a\atop\alb}\sum_{q_{1,2,3}}B_{q_1}^{a\alpha}
%B_{q_2}^{a\alpha}B_{q_3}^{a\beta}B_{q_4}^{a\beta}\non
%&&+4T\ln\left[\frac{\Gamma(\frac{3}{2})}{2\pi^{3/2}}\right]
%-\sum_\alphaH^\alpha\cdot\left(\sum_aB_o^{a\alpha}\right)
\end{equation}
%where $q_4=-(q_1+q_2+q_3)$.  
Consider diagonalization of the $3n\times 3n$ matrix, $J^{a\alpha
b\beta}_{\bf q}$. Denote the eigenvalues by $\lambda_{\bf q}^i$ and
eigenvectors by ${\mathbf U}^{i}_{\bf q}$ for $i=1\ldots 3n$.  Expanding the
order parameters in the basis of eigenvectors
\begin{equation}
B_{\bf q}^{a\alpha}=\sum_{i} U_{\bf q}^{a,\alpha i}\phi_{\bf q}^{i}
\end{equation}
one obtains
\begin{equation}
f^{(2)}=\frac{1}{2}\sum_{\bf q}\sum_{i}\left(3T+\lambda_{\bf q}^{i}\right)\phi_{\bf q}^{i}\phi_{-{\bf q}}^{i}.
%&&+\frac{9T}{20}
%\sum_{q_{1,2,3}}\sum_{a\atop\alb}\sum_{ijkl}\phi_{q_1}^i \phi_{q_2}^j
%\phi_{q_3}^k \phi_{q_4}^l\non
%&&\times U_{q_1}^{a\alpha i} U_{q_2}^{a\alpha j}
%U_{q_3}^{a\beta k} U_{q_4}^{a\beta l}\non
%&&+4T\ln\left[\frac{\Gamma(\frac{3}{2})}{2\pi^{3/2}}\right]
%-\sum_\alpha H^\alpha\sum_{a,i}U_o^{a\alpha i}\phi_o^i.
\end{equation}
To find the minimum of the quadratic term, it is necessary to determine
the minimum $\lambda_{\bf q}^{i}$ for all $i$ and
${\bf q}$. 

Specializing to the pyrochlore lattice, we follow previous workers in
using a rhombohedral unit cell containing four magnetic ions
($3n=12$).\cite{reimers91,raju99} Basis vectors are
$\frac{a}{2}(\hat{\jmath} + \hat{k})$, $\frac{a}{2}(\hat{\imath} +
\hat{k})$, and $\frac{a}{2}(\hat{\imath} + \hat{\jmath})$, where $a$
is the edge length of the cubic unit cell.  Also, the four magnetic
ions within the unit cell are located at $(x,y,z)= (0,0,0)$,
$(0,\frac{1}{4},\frac{1}{4})$, $(\frac{1}{4},0,\frac{1}{4})$, and
$(\frac{1}{4},\frac{1}{4},0)$.

The instability of the paramagnetic phase is analyzed by considering
the sign of $(3{\mathrm T} +\lambda_{\bf q}^i)$.  In the paramagnetic phase,
all are positive.  At the transition, the smallest becomes negative.
Using the ratio $J_{dd}/J_{ex}=0.2$, Raju \textit{et al.}  showed that
such an approach does not completely determine the ordering
pattern.\cite{raju99} At $T_c$, a star of modes simultaneously becomes
unstable.  For the same value of $J_{dd}/J_{ex}$ , we recover these
results.  By contrast, for values of $J_{dd}/J_{ex}>J_c\simeq 5.7$, we
find that a discrete set of isolated modes (related by the lattice
symmetry) become unstable.

\begin{figure}
\begin{center}
\resizebox{!}{2.5in}{\includegraphics{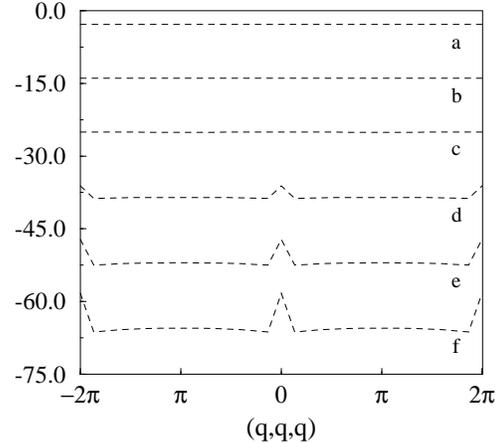}}
\caption{Minimal eigenvalues along the $(111)$ line in $q$-space for
various values of the dipole-dipole interaction strength.  Curve $a$
corresponds to $J_{dd}/J_{ex}=0.2$ as in Ref. 12.
Curves $b$ through $f$ have $J_{dd}/J_{ex}=2.92$, $5.63$, $8.35$, $11.06$, and
$13.78$, respectively.\cite{lrgJddnote}\label{jddlarge}}
\end{center}
\end{figure}

To illustrate this point, we plot in Figure \ref{jddlarge} the minimum
eigenvalue of the $J_{\bf q}$ matrix as a function of ${\bf q}$ for various
values of $J=J_{dd}/J_{ex}$ ranging from that used by
Raju \textit{et al.}\cite{raju99} (based on the measured values of
$|\theta_{CW}|$ and the lattice spacing and the calculated value of
the magnetic moment of ${\mathrm Gd}^{3+}$) to a much larger value.
For $J<J_c$, the minimum eigenvalue is independent of $\bf q$ along
the $(111)$ direction.  For $J>J_c$, there are isolated minima located
close to (but not at) ${\bf q}={\bf 0}$.  These are the individual
modes that become unstable at $T_c$ in the case $J>J_c$.

\begin{figure}
\begin{center}
\leavevmode
\hbox{%
\epsfysize=1in
\epsffile{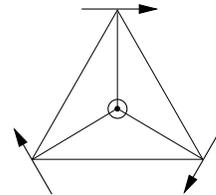}}
\caption{An illustration of the eigenvectors for the soft modes along
the star of the (111) directions in {\bf q}-space.  The vector {\bf q}
is perpendicular to the base of the tetrahedron and is shown coming
out of the page. The three spins are coplanar, have equal magnitude,
and are tangent to the cirle which circumscribes the base of the
tetrahedron.  The spin on the fourth site has zero magnitude.
\label{mode}}
\end{center}
\end{figure}

\begin{figure}
\begin{center}
\leavevmode
\hbox{%
\epsfysize=2in
\epsffile{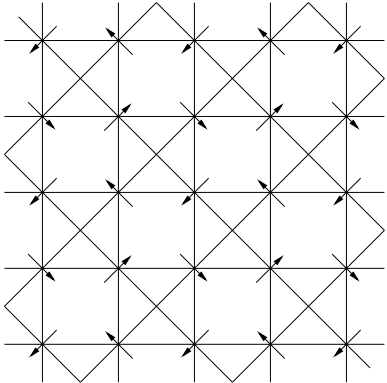}}
\caption{A projection of our ${\bf q}={\bf 0}$ ordering pattern onto the
$xy$-plane.  The spins are coplanar and form a 4-sublattice N\'eel state.}
\label{pyrogs}
\end{center}
\end{figure}

We now investigate the ordering pattern below $T_c$ for $J<J_c$.  The
degeneracy of the soft modes is lifted when we include contributions
to the free energy from the fourth order term.  At temperatures below
$T_c$, the order parameters acquire finite magnitude and one is faced
with the problem of simultaneously minimizing the quadratic and
quartic parts of the free energy.  Consider first the quartic term in
isolation.  In terms of the real space order parameters,
$B_{la}^\alpha$, this is
\begin{equation}
F^{(4)}=\frac{9T}{20}\sum_{la,\alb}B_{la}^{\alpha}B_{la}^{\alpha}B_{la}^{\beta}
B_{la}^\beta .
\end{equation}
Its value depends on the magnitude and direction of ${\mathbf
B}_{la}$.  For fixed magnitude of
$\Sigma_{la\alpha}(B_{la}^\alpha)^2$, the quartic term is minimized by
a state with all $|{\mathbf B}_{la}|$ equal.  Fortunately, and
apparently fortuitously, a state satisfying this condition can be
constructed which also minimizes the quadratic term.  In detail, we
proceed as follows. The eigenvector associated with one of the modes
that becomes unstable at $T_c$ is illustrated in
Fig.\,\ref{mode}. Taking the most general combination of this
eigenvector and its three partners, and imposing the condition that
the values of $|{\bf B}_{la}|$ are the same at four sites of a
tetrahedron, we generate the configurations of Fig.\,\ref{gnds} and no
others. Tiling the lattice with these configurations, we obtain only
states with ordering wavevector ${\bf q} = {\bf 0}$. We conclude that
the ordering pattern for $T$ just below $T_c$ is as shown in
Fig.\,\ref{pyrogs}.

\section{Static distortions away from the ordered state}
\begin{figure}
\begin{center}
%\leavevmode
%\hbox{%
%\epsfysize=2.2in 
%\epsffile{varJfig.eps}}
\resizebox{!}{2.5in}{\includegraphics{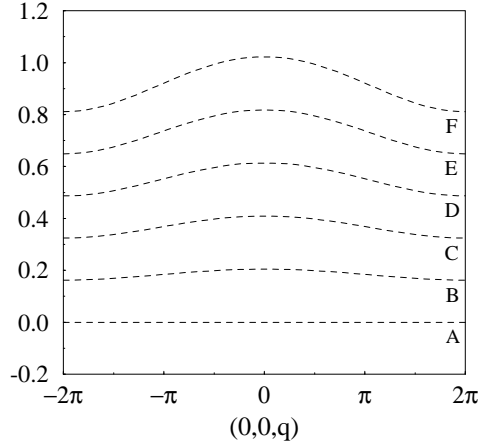}}
\caption{The minimum eigenvalues for distortions away from the ground
state in the $(001)$ direction in $q$-space.  The various curves
correspond to different values of the relative strength of the
dipole-dipole interaction, $J_{dd}/J_{ex}$.  Curves $A$ through $E$ have
$J_{dd}/J_{ex}=0.0$, $0.04$, $0.08$, $0.12$, and $0.16$, respectively.  Curve
$F$ has $J_{dd}/J_{ex}=0.2$ as in Ref. 12.
\label{quaddis}}
\end{center}
\end{figure}

Finally, we investigate whether a ground state with this ordering
pattern is stable against thermal fluctuations at low temperatures.
To describe this calculation using the notation introduced above, we impose
the constraint $|{\mathbf B}_{la}|=1$ for every $l$ and $a$.  We
denote the components of ${\bf B}_{la}$ in the ground state with
the ordering pattern shown in Figure \ref{pyrogs} by $B^0_{la\alpha}$.  We
expand 
\begin{equation}
B_{la\alpha}=B_{la\alpha}^0+\delta
B_{la\alpha}-\frac{1}{2}B_{la\alpha}^0|\delta  {\bf
B}_{la}|^2+{\mathcal O}({\mathbf \delta B}^4).
\end{equation}
Since $\delta {\bf B}_{la}$ (when small) is orthogonal to ${\bf B}_{la}$,
it has only two independent components. 
%Specifically, we take 
%${\mathbf \delta B}_{la}$ to have the form in Cartesian
%coordinates
%\begin{eqnarray}
%\boldsymbol{\delta}{\bf B}_{l1}&=&(d_{l1},d_{l1},e_{l1})\non
%\boldsymbol{\delta}{\bf B}_{l2}&=&(d_{l2},-d_{l2},e_{l2})\non
%\boldsymbol{\delta}{\bf B}_{l3}&=&(d_{l3},-d_{l3},e_{l3})\non
%\boldsymbol{\delta}{\bf B}_{l4}&=&(d_{l4},d_{l4},e_{l4})
%\end{eqnarray}
%where site labels within the basis, $a=1,2,3$, and $4$, correspond to
%the sites at $(x,y,z)=(0,0,0),(0,\frac{1}{4},\frac{1}{4}),
%(\frac{1}{4},0,\frac{1}{4})$, and $(\frac{1}{4},\frac{1}{4},0)$.  
We expand the energy to quadratic order in the distortion variables
obtaining
\begin{eqnarray}
{\mathcal
H}=E_0&+&\frac{1}{2}\sum_{lmab,\alb}(J_{lmab}^{\alb})\times(\delta
B_{la\alpha}\delta
B_{mb\beta}\non 
&&-\frac{1}{2}(|\delta{\mathbf B}_{la}|^2+|\delta{\mathbf B}_{mb}|^2){\hat
B}_{la\alpha}^0{\hat B}_{mb\beta}^0)
\end{eqnarray}
where $E_0$ is the energy of the ground state.  We then diagonalize
the quadratic term, which involves an $8\times 8$ matrix for each
wavevector, ${\mathbf q}$.  We find the minimum eigenvalues of this
matrix as a function of ${\mathbf q}$.  As expected, we recover a flat
branch of zero-energy modes when $J_{dd}=0.0$.\cite{reimers91} For
$J_{dd} > 0.0$, we find that all fluctuations away from an ordered
state are associated with a positive energy cost (even in the long
wavelength limit, since dipolar interactions break global rotational
symmetry).  As an illustration, we plot the minimum at each ${\mathbf
q}$ along the $(001)$ direction in $q$-space for various values of the
dipole-dipole interaction in Figure \ref{quaddis}.  The energetic
barriers around our proposed ground state at ${\mathrm T}=0$ mean that
there is not a degenerate, connected manifold.  The ordering obtained
is due to energetic selection and does not occur via an
order-by-disorder mechanism.

\section*{summary}
We have considered the influence of dipolar interactions in a nearest
neighbor Heisenberg model of a geometrically frustrated system,
reproducing and extending earlier work by Raju {\it
et. al}.\cite{raju99} In particular, we determine the ordering
pattern, using a mean field treatment and by examining the stability
of an ordered state to fluctuations. We believe that the ordering we
find should be that associated with the transition which is observed
experimentally at around 1K in Gd$_2$Ti$_2$O$_7$.\cite{raju99} This
ordered state, shown in Figure \ref{pyrogs}, is a four sub-lattice
N\'{e}el state.

It will be interesting to compare these theoretical predictions with
the results of neutron scattering from $^{160}$Gd$_2$Ti$_2$O$_7$, currently
in progress.\cite{bram99}

\section*{acknowledgments}
We thank M. J. P. Gingras for very helpful comments 
on a draft version of this manuscript.
S. E. P. would like to thank the Rhodes Trust for financial support.
J. T. C. acknowledges support under EPSRC Grant Number GR/J78327.
%\nocite{*}


\begin{thebibliography}{10}
\bibitem{lrgJddnote}
Since the dipolar interactions are long-range, it is important to ensure that
  finite size effects are not substantial. These results are for a system
  containing 62,500 spins. We find that the results are the same for larger
  system sizes.
\end{thebibliography}
\end{document}